\documentclass{ws-procs9x6-cpt16}
\begin{document}

\newcommand{\refeq}[1]{(\ref{#1})}
\def\etal {{\it et al.}}

\def\Hbar{\mbox{$\overline{\mathrm H}$}}
\def\s{$\sigma_1$}
\def\p{$\pi_1$}
\def\Bext{\mbox{$B_{\mathrm{ext}}$}}
\def\Bosc{\mbox{$B_{\mathrm{osc}}$}}

\def\nr{\rm NR}
\def\nrtemplate#1#2#3{#1^{\nr#3}_{#2}}

\def\anrf#1#2{\nrtemplate{{a_{#1}}}{#2}{}}
\def\cnrf#1#2{\nrtemplate{{c_{#1}}}{#2}{}}
\def\gzBnrf#1#2{\nrtemplate{{g_{#1}}}{#2}{(0B)}}
\def\goBnrf#1#2{\nrtemplate{{g_{#1}}}{#2}{(1B)}}
\def\goEnrf#1#2{\nrtemplate{{g_{#1}}}{#2}{(1E)}}
\def\HzBnrf#1#2{\nrtemplate{{H_{#1}}}{#2}{(0B)}}
\def\HoBnrf#1#2{\nrtemplate{{H_{#1}}}{#2}{(1B)}}
\def\HoEnrf#1#2{\nrtemplate{{H_{#1}}}{#2}{(1E)}}

\title{Prospects of In-Flight Hyperfine Spectroscopy of\\
(Anti)Hydrogen for Tests of CPT Symmetry}

\author{E.\ Widmann}

\address{Stefan Meyer Institute for Subatomic Physics,
1090 Vienna, Austria}

\author{On behalf of the ASACUSA-CUSP Collaboration}

\begin{abstract}
The ground-state hyperfine splitting of antihydrogen promises one of the most sensitive tests of CPT symmetry. The ASACUSA collaboration is pursuing a measurement of this splitting in a Rabi-type experiment using a polarized beam from a CUSP magnet at the Antiproton Decelerator of CERN. With the initial intention of characterizing the Rabi apparatus, a polarized source of cold hydrogen was built and the $\sigma_1$ transition of hydrogen was measured to a few ppb precision. A measurement of the $\pi_1$ transition is being prepared. The availability of this beam opens the possibility to perform first measurements of some coefficients within the nonminimal Standard-Model Extension.
\end{abstract}

\bodymatter

\section{Introduction}

The spectroscopy of antihydrogen offers some of the most sensitive tests of CPT symmetry, because its CPT counterpart hydrogen is one of the most precisely known atoms. The ASACUSA collaboration has proposed a measurement of the ground-state hyperfine splitting (GS-HFS) of antihydrogen in a Rabi-type atomic beam setup\cite{Widmann:2013qy} using antihydrogen generated in a magnetic CUSP configuration.\cite{Mohri:2003wu} A first beam of antihydrogen atoms produced this way was recently observed in a region free from stray magnetic fields,\cite{Kuroda:2014fk} and efforts are currently under way at the CERN Antiproton Decelerator to increase the rate of antihydrogen atoms and to determine their properties like quantum state and velocity.

The central part of the Rabi apparatus consists of a microwave cavity for flipping the antihydrogen spin and a superconducting sextupole magnet for spin analysis.\cite{Widmann:2013qy} In order to verify the obtainable resolution of this setup, a source of cold and polarized hydrogen atoms was built at the Stefan Meyer Institute and combined with a Q-mass spectrometer to measure the GS-HFS of hydrogen.\cite{DiermaierEtAl2015}

\section{Hyperfine splitting and the SME}\label{EBW:sec1}

In a classical atomic beam resonance experiment, the transition is induced in a constant external magnetic field \Bext. In this case, two types of transitions can be observed
between the $F=0$ singlet and $F=1$ triplet states:
the $\Delta M_F=0$ so-called \s\ transition  $(F,M_F)=(1,0) \rightarrow (0,0)$, and the $\Delta M_F=1$ \p\ transition from the highest lying state ($(1,1)$ for H and ($1,-1$) for \Hbar) to $(0,0)$. Within the Standard-Model Extension, only transitions with $\Delta M_F \ne 0$ are sensitive to Lorentz or CPT violation.

In the minimal SME, the energy shift of HFS states is determined by the coefficients $\tilde{b}_X^w=b_X^w - d^w_{X0}m_w - H^w_{YZ}$,
$\tilde{b}_Y^w$, and $\tilde{b}_Z^w$,\cite{Bluhm:1999vq} where $X,Y,Z$ denote the projection of the tensor couplings on a fixed inertial frame, $w$ stands for electron/positron or proton/aniproton, and $m_w$ is the respective mass. CPT violation leads to the coefficients $d$ and $H$ reversing sign for antihydrogen. $\tilde{b}_{X,Y}^e$ and $\tilde{b}_{X,Y}^p$ have been constrained by measurements using a hydrogen maser\cite{HumphreyEtAl2003} to values of the order $10^{-27}$ GeV.

The nonminimal SME offers more opportunities for hyperfine measurements with (anti)hydrogen.\cite{KosteleckyVargas2015} Again only the \p\ transition experiences a shift that depends on the nonrelativistic spherical coefficients
$\gzBnrf{w}{(2q)10}$,
$\goBnrf{w}{(2q)10}$ (CPT odd) and $\HzBnrf{w}{(2q)10}$, $\HoBnrf{w}{(2q)10}$ (CPT even). The time-dependent coefficients are already bound by the maser measurements mentioned above,\cite{KosteleckyVargas2015} but those which depend instead on the orientation of \Bext\ can be determined for the first time by measurements using the ASACUSA hydrogen beam. This apparatus can in addition be used for deuterium spectroscopy by changing the spin-flip cavity and maybe the detection scheme to suppress background from residual H$_2$ gas.  Deuterium is of special interest, not only because it offers in addition access to coefficients in the neutron sector but also because the Fermi motion of the proton in the nucleus permits a billionfold improvement for some coefficients compared to hydrogen.\cite{KosteleckyVargas2015}

\section{Hydrogen GS-HFS measurement in a beam}

The hydrogen beam setup consists of three parts (cf.\ Fig.\ 1):
(i) a source comprising an RF discharge to produce atomic hydrogen from molecular H$_2$ gas, a PTFE tube cooled to cryogenic temperature,  two permanent sextupole magnets serving as polarizer and velocity selector, and a tuning fork chopper for background suppression;
(ii) the HFS spectrometer, consisting of a microwave cavity and a superconducting sextupole; and (iii) a quadrupole mass spectrometer to detect atomic hydrogen.

\begin{figure}[t]
\begin{center}
\includegraphics[width=0.8\hsize]{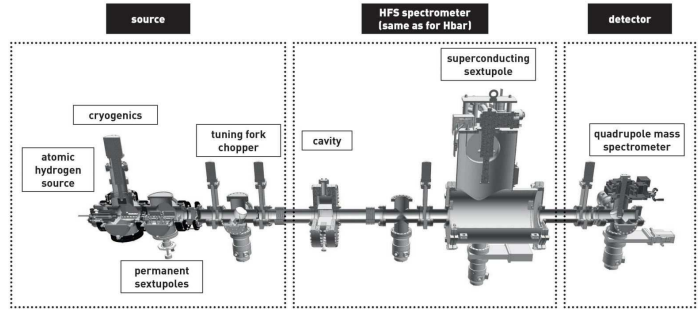}
\caption{Overview of the hydrogen beam setup. For detail see text.}
\end{center}
\label{EBW:figHbeam}
\end{figure}

In order to reproduce the expected beam temperature of antihydrogen generated in our CUSP magnet, the hydrogen atoms are cooled to $\sim50$~K corresponding to a velocity of $v_{\mathrm{H}}\simeq 1$ km/s. The velocity selection by two permanent sextupole magnets leads to a gaussian velocity distribution with about 100 m/s (1 $\sigma$) velocity spread. Currently the external magnetic field \Bext\ (typical strength of  few gauss) is aligned parallel to the oscillating magnetic field \Bosc, thus exciting the \s\ transition at around $1.42$ GHz. The resulting resonance curve (Fig.\ 2) shows a double dip structure which originates from the strip-line cavity used that has a spatially varying oscillating magnetic field in the beam direction. The line shape is well understood by solving the von Neumann equation of the density-matrix formalism  for a fixed velocity and overlaying several solutions to obtain a velocity distribution. The center of the resonance was measured at several values of \Bext\ and the zero-field values were extracted using the  dependence obtained from the Breit-Rabi formula. In one extrapolation sequence, the zero-field hyperfine frequency can be obtained with an accuracy of order 10 ppb, which agrees with the value obtained in a maser within one standard deviation. Ten different extrapolations have been recorded, so that we expect to be able to determine the zero-field value to a few ppb precision.

\begin{figure}[t]
\begin{center}
\includegraphics[width=\hsize]{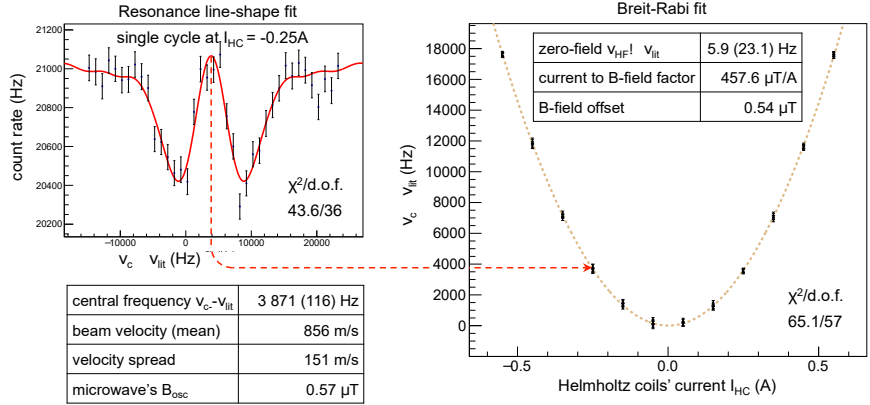}
\caption{Left: Resonance scan and fit of a line shape function described in the text. The table on the lower left gives the fit parameters. Right: extrapolation of resonance frequency obtained at several Helmholtz coil currents $I_\mathrm{HC}$ corresponding to \Bext\ values to zero field.}
\end{center}
\label{EBW:figZeroField}
\end{figure}

\section{Outlook}

As a next step for hydrogen, the measurement of the \p\ transition is planned, which is much more sensitive to magnetic-field inhomogeneities. An improved magnetic shielding and correction coils for the Helmholtz coils are being constructed. The main limitation in accuracy of our apparatus comes from the transit time broadening of the atoms flying through the cavity, which at 1 km/s velocity leads to a line width of 12 kHz. The line center has been determined to a few Hz, splitting the line by more than a factor 1000. A further increase in precision can be obtained by reducing the velocity of hydrogen atoms to which the line width is  proportional.

The result shows that the resonance method works. Together with the observation of an antihydrogen beam in a field-free region all prerequisites are now available for a measurement with antihydrogen.
Due to the much lower statistics, a precision in the order of ppm is expected in this case.\cite{KolbingerEtAl2015}

\section*{Acknowledgments}
This work was funded by the European Research Council (grant number 291242), the Austrian Ministry of
Science and Research, and the Austrian Science Fund (FWF): DKPI(W 1252).

\end{document}